\documentclass[10pt,a4paper,english,preprint,aps]{revtex4-1}
\usepackage{mathptmx}

\usepackage[T1]{fontenc}
\usepackage[latin9]{inputenc}
\usepackage{color}
\usepackage{babel}
\usepackage{float}
\usepackage{textcomp}
\usepackage{amstext}
\usepackage{graphicx}
\usepackage{subscript}
\usepackage[unicode=true,pdfusetitle,
 bookmarks=true,bookmarksnumbered=false,bookmarksopen=false,
 breaklinks=false,pdfborder={0 0 1},backref=false,colorlinks=false]
 {hyperref}
\hypersetup{pdftitle={Characterization of the long-term dimensional stability of a NEXCERA block using the optical resonator technique},
 pdfauthor={Chang Jian Kwong,  Michael G. Hansen, Jun Sugawara, and  Stephan Schiller}}

\makeatletter

\pdfpageheight\paperheight
\pdfpagewidth\paperwidth

\usepackage{environ}
\usepackage{linkToId}

\makeatother

\begin{document}

\title{{\normalsize{}Characterization of the long-term dimensional stability
of a NEXCERA block using the optical resonator technique}}

\author{{\normalsize{}\linkToOrcidNameIcon{Chang Jian Kwong}{https://orcid.org/0000-0001-5328-5853}}\textsuperscript{1}{\normalsize{},
\linkToOrcidNameIcon{Michael G. Hansen}{https://orcid.org/0000-0001-5180-2511}}\textsuperscript{1}{\normalsize{},
Jun Sugawara}\textsuperscript{2}{\normalsize{}, and \linkToOrcidNameIcon{Stephan Schiller}{https://orcid.org/0000-0002-0797-8648}}\textsuperscript{1}}

\address{\noindent 1. Institut für Experimentalphysik, Heinrich-Heine-Universität
Düsseldorf, Düsseldorf, Germany}

\address{\noindent 2. Technical Management Department, Krosaki Harima Corporation,
Kitakyushu-City 806-8586, Japan }

\address{Email: \href{mailto:Step.Schiller@hhu.de}{Step.Schiller@hhu.de}}
\begin{abstract}
{\normalsize{}NEXCERA is a machinable and highly polishable ceramic
with attractive properties for use in precision instrument}\textcolor{black}{\normalsize{}s,
in particular because its coefficient of thermal expansion exhibits
a zero crossing at room temperature. We performed an accurate measurement
of the long-term drift of the length of a 12~cm long NEXCERA block
by using it as a spacer of a high-finesse optical cavity. At room
temperature, we found a fractional length drift rate $L^{-1}d\Delta L/dt=-1.74\times10^{-8}$~$\mathrm{yr}^{-1}$. }{\normalsize \par}
\end{abstract}
\maketitle

\section{{\normalsize{}Introduction}}

Materials with ultra-low thermal expansion and high long-term dimensional
stability are highly desirable in the field of precision instruments
and metrology. Two well-known examples are ULE (ultra-low expansion
glass) from Corning, Inc. and Zerodur from Schott AG. Both ULE and
Zerodur have a small coefficient of thermal expansion $\alpha$, quoted
as $0\pm0.03\times10^{-6}\,$/K in the range 278~K - 308~K \citep{Corning2},
and $0\pm0.006\times10^{-6}\,$/K in the range 273~K - 323~K \citep{Schott2},
respectively. The long-term dimensional stability of materials has
been the subject of many studies, e.g. \citep{Berthold1977,Jacobs1989}.
For ULE, values of the fractional length drift rate $L^{-1}d\Delta L/dt\approx-2.5\times10^{-8}$~$\mathrm{yr}^{-1}$
have been reported \citep{Zhu1992}, as well as significantly lower
ones, $-0.4\times10^{-8}$~$\mathrm{yr}^{-1}$ \citep{Marmet1997}.
Zerodur has a relatively large dimensional instability, $-0.1\times10^{-6}$
to $-0.2\times10^{-6}$~$\mathrm{yr}^{-1}$, \citep{Takahashi2010,Zhu1992},
making it less suitable for applications requiring extreme long-term
stability.

A relatively recent ultra-low thermal expansion ceramic, NEXCERA,
developed by Krosaki Harima (Japan), is also a material of interest
in instrumentation. NEXCERA can be produced with a thermal expansion
coefficient $|\alpha|<0.03\times10^{-6}$~/K at $23\,^{\circ}$C.
The Young's modulus, 140~GPa, is higher than that of both ULE (67.6~GPa)
and Zerodur (90.3~GPa). NEXCERA's bulk density, 2.58~$\mathrm{g/cm}^{3}$,
is comparable to Zerodur (2.53~$\mathrm{g/cm}^{3}$) but moderately
higher than ULE's (2.21~$\mathrm{g/cm}^{3}$). 

The first characterization of the dimensional stability of NEXCERA
was recently reported by Takahashi \citep{Takahashi2012}. Length
measurements of line scales made from NEXCERA showed that the ceramic
has good long-term dimensional stability. The fractional length change
$\delta L/L$ over a time interval of 13~months was determined as
$(1.1\pm1.1)\times10^{-8}$ ($2\sigma$ error) \citep{Takahashi2012},
which is consistent with zero length change. On fundamental grounds
and for extremely demanding applications it is desirable to determine
the dimensional stability of NEXCERA more precisely. This was the
purpose of the present work. 

An extremely sensitive method for determining the long-term dimensional
stability of a material consists of manufacturing the material into
a spacer for an optical cavity. One of the mode frequencies of such
a cavity is then measured repeatedly against an atomic frequency standard. 

For ULE, such measurements have been done in many laboratories world-wide.
The lowest long-term linear drift rates $L^{-1}d\Delta L/dt\approx5\times10^{-17}\,{\rm s}^{-1}$
($0.15\times10^{-8}$~$\mathrm{yr}^{-1}$) have recently been found
\citep{Hafner2015,Keupp2005}. Other measurements of ULE long-term
linear drift include $L^{-1}d\Delta L/dt=2.04\times10^{-16}\,{\rm s}^{-1}$
($0.64\times10^{-8}$~$\mathrm{yr}^{-1}$) \citep{Alnis2008} and
$L^{-1}d\Delta L/dt=1.45\times10^{-16}\,{\rm s}^{-1}$ ($0.44\times10^{-8}$~$\mathrm{yr}^{-1}$)
\citep{Keller2014}. A measurement of the long-term linear drift rate
of a Zerodur optical resonator has shown $L^{-1}d\Delta L/dt=3.05\times10^{-15}{\rm s}^{-1}$
($9\times10^{-8}$~$\mathrm{yr}^{-1}$) \citep{Keupp2005}, measured
over 2 years.

A NEXCERA optical resonator has been realised by Hosaka \emph{et.
al.} \citep{Hosaka2013}. It consisted of a 75~mm long cylindrical
NEXCERA spacer and a pair of ULE mirror substrates. The temperature
at which the coefficient of thermal expansion of the resonator is
zero (zero-CTE temperature), was found to be at $T_{0}=16.4\pm0.1\,^{\circ}\mathrm{C}$
\citep{Hosaka2013}. As an upper limit of the drift rate, $L^{-1}|d\Delta L/dt|<1.2\times10^{-7}$~$\mathrm{yr}^{-1}$
was given. This corresponds to a drift of the optical frequency of
1~Hz/s at 1.064~\textmu m wavelength. 

In this work, we present the first accurate measurement of the long-term
drift of a NEXCERA\textcolor{black}{{} N118C opt}ical resonator, which
is found to be non-zero. 

This paper is structured as follows: the experimental setup is presented
in Sec.~\ref{sec:Experimental-Setup}. The characterization of zero-CTE
temperature and the characterization of the long-term drift of the
NEXCERA resonator are discussed in Sec.~\ref{sec:Results-and-Discussion}.
We draw conclusions in Sec.~\ref{sec:Conclusion}.

\section{{\normalsize{}Experimental Setup\label{sec:Experimental-Setup}}}

\begin{figure}[H]
\begin{centering}
\includegraphics[scale=0.16]{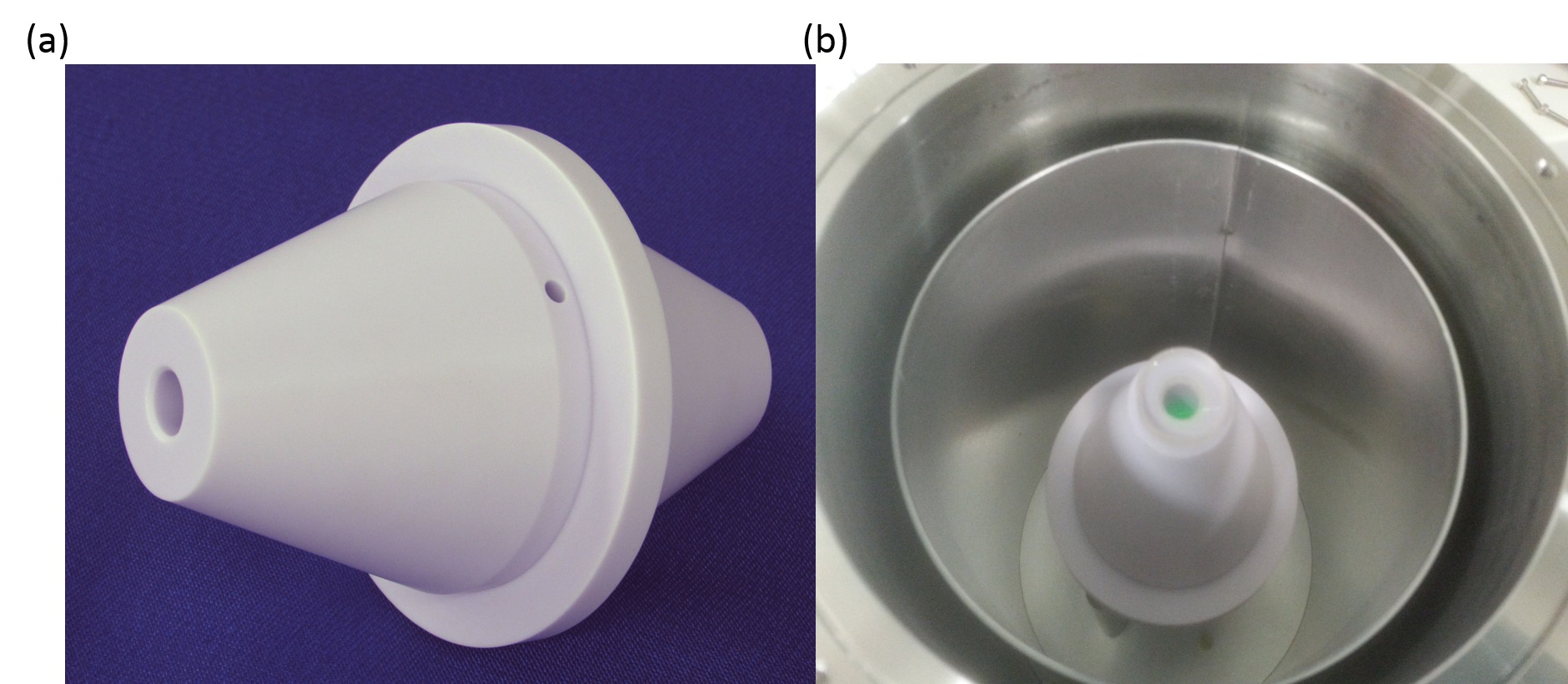}
\par\end{centering}
\caption{(a) NEXCERA spacer. (b) NEXCERA optical resonator within its vacuum
chamber.\label{fig:The-NEXCERA-spacer.}}
\end{figure}

The NEXCERA sample we used (sample number N118C) was sintered in December
2015 and machined in March 2016. The resonator consists of a NEXCERA
spacer of biconical shape with a central rim, see Fig.~\ref{fig:The-NEXCERA-spacer.}~(a).
The length is 120~mm, the conical angle is $20\,^{\circ}$. The diameter
at the endfaces and the rim are 30~mm and 90~mm respectively. A
10~mm diameter center bore allows the confined light to propagate
in vacuum. Evacuation occurs via a 4~mm diameter transverse pumping
hole.

\begin{figure}[H]
\centering{}\includegraphics[scale=0.3]{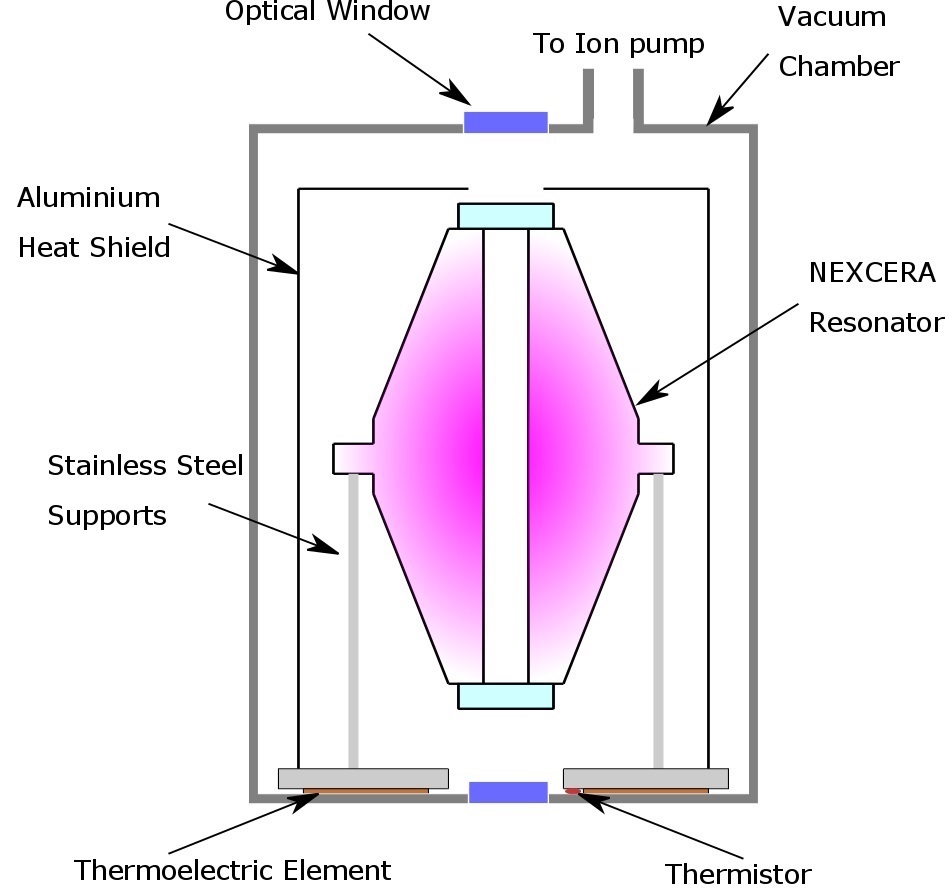}\caption{\label{fig:Setup}Schematic diagram of the NEXCERA resonator mounted
within its vacuum chamber.}
\end{figure}

Two ULE concave mirror substrates with 0.5~m radius of curvature,
having a high reflectivity coating for 1064~nm, are optically contacted
to the cavity spacer in-house. The contacting was done on 21~December
2016. The resonator linewidth was measured to be 20~kHz (FWHM) (finesse
of $62\times10^{3}$). Fig.~\ref{fig:The-NEXCERA-spacer.}~(b) and
\ref{fig:Setup} illustrate the mounting of the NEXCERA resonator
within a custom-made vacuum chamber. The NEXCERA cavity is placed
vertically on three stainless steel supports with viton cylinders
(approximately 5~mm long and 1.5 mm in diameter) placed in-between
the supports and the resonator. These provide thermal insulation and
mechanical damping. The steel supports are fixed to a base plate.
Thermoelectric elements glued onto the bottom side of the base plate,
together with a thermistor for temperature measurement, allow active
stabilization of the temperature by a PID controller. The resonator
and base plate are placed inside a polished aluminium heat shield
which provides for further thermal insulation by reflecting radiative
heat from the environment.

\begin{figure}[H]
\begin{centering}
\includegraphics[scale=0.3]{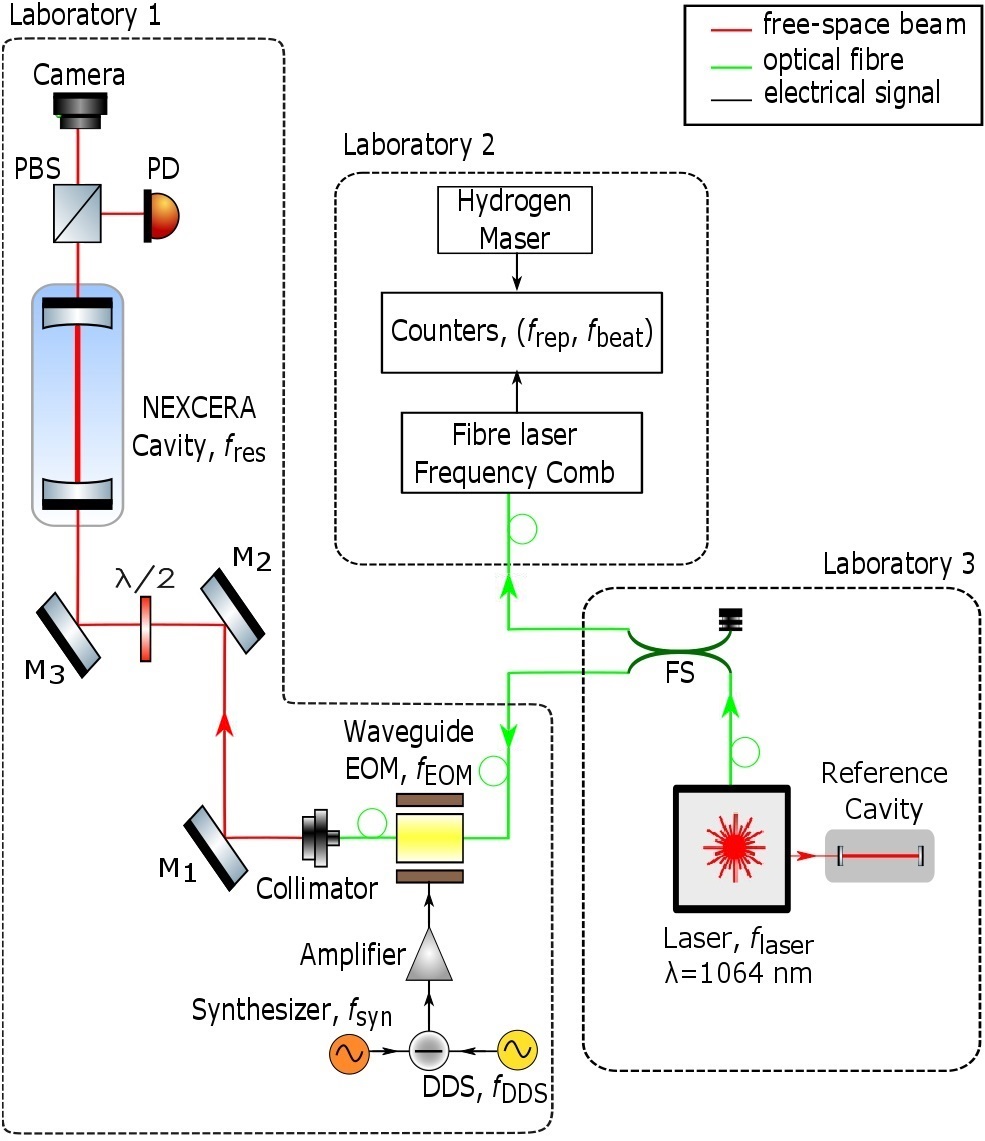}
\par\end{centering}
\caption{Schematic diagram of the optical setup for the characterization of
the resonator's long-term frequency drift.\label{fig:The-optical-setup}}
\end{figure}

Fig.~\ref{fig:The-optical-setup} shows the optical setup. The frequency
of a particular TEM$_{00}$ mode of the resonator is interrogated
by a Nd:YAG reference laser (frequency $f_{{\rm laser}}$) which is
stabilized in frequency to an independent ultra-stable ULE reference
cavity \citep{Chen2014}. With a fibre splitter, the light from the
reference laser is split into two arms. One arm directs the light
to an erbium-doped fibre-based frequency comb referenced to a hydrogen
maser, while the other arm sends the light to the NEXCERA cavity.
The laser's frequency, as measured with the frequency comb, is determined
from
\begin{center}
\begin{equation}
f_{\mathrm{laser}}=n\,f_{\mathrm{rep}}\pm f_{\mathrm{beat}}\pm f_{\mathrm{\mathrm{offset}}},\,\label{eq:frequencyComb}
\end{equation}
\par\end{center}

\noindent where $f_{\mathrm{rep}}$ is the frequency comb repetition
rate, $f_{\mathrm{beat}}$ is the frequency of the beat note between
the selected mode of the frequency comb and the Nd:YAG laser, \emph{f}\textsubscript{offset}
is the comb carrier envelope offset frequency and $n$ is the number
of the frequency comb's mode. In our apparatus, $f_{\mathrm{rep}}\simeq250\,$MHz,
\emph{f}\textsubscript{offset}$=20$~MHz, $f_{\mathrm{beat}}=50$~MHz,
$n\approx1.126\times10^{6}$. The sign of \emph{f}\textsubscript{offset}
is determined by a particular setting of the locking electronics.
The frequency of the laser is first coarsely measured with a wavemeter
and the measured value, $f_{\mathrm{laser,w}}$ is used as a starting
point for the more precise measurement with the frequency comb. The
correct sign for $f_{\text{beat}}$ in Eqn.~(\ref{eq:frequencyComb})
is chosen so that the measured value of $f_{\mathrm{laser}}$ is closest
to $f_{\mathrm{laser,w}}$. With $f_{\mathrm{beat}}$ fixed at 50~MHz,
the change in the frequency comb's repetition rate due to drifts of
the reference laser $f_{\mathrm{laser}}$ can be determined with high
precision. $f_{\mathrm{laser}}$, as measured by the comb, is then
averaged over the 15~s duration of each frequency scan, resulting
in the average laser frequeny, $f_{\mathrm{laser,av}}$ corresponding
to each frequency scan. The averaging time of 15~s is chosen because
it corresponds to the duration of one frequency scan of the NEXCERA
cavity as explained below. Since the drift of the frequency-stabilized
Nd:YAG laser is small ($df_{\mathrm{laser}}/dt=$0.055~Hz/s), its
frequency does not change significantly during the averaging time.

The other part of the laser light that is transmitted through the
NEXCERA cavity is split by a polarising beam splitter (PBS) and sent
to a camera and a photodetector (PD) for measurement. A half-wave
plate ($\lambda/2$) is used for maximisation of the photodetector
signal. Using a waveguide electro-optic modulator (EOM), we generate
two sidebands on the laser wave before the light is sent to the NEXCERA
cavity. Due to frequency bandwidth limitations of the RF sources used,
the RF signal driving the EOM (frequency $f_{\mathrm{EOM}}$) is produced
by mixing an RF signal having constant frequency ($f_{\mathrm{syn}}=607.663$~MHz)
with an RF signal (frequency $f_{\mathrm{DDS}}$$\approx60\,$MHz)
from a direct digital synthesizer (DDS), $f_{\mathrm{EOM}}=$$f_{\mathrm{syn}}-f_{\mathrm{DDS}}$.
By varying $f_{\mathrm{DDS}}$, one of the laser sidebands can be
scanned across the resonator's mode frequency. Thus, the resonator's
frequency can be determined, as well as its linewidth. The synthesizer
and the DDS are both referenced to the maser.

The wave exiting from the EOM is coupled into a TEM\textsubscript{00}
mode of the resonator. The transmitted light is split by a polarising
beam splitter (PBS) and sent to a camera and a photodetector (PD)
for measurement. A half-wave plate ($\lambda/2$) is used for maximization
of the photodetector signal.

A computer controls the DDS and acquires the photodetector (PD) signal.
The employed DDS frequency scan has a span of 100~kHz and a step
size of 1~kHz with a dwell time of 150~ms. This amounts to 15~s
duration per line scan. Such frequency scans are typically repeated
for approximately 1 hour in the measurement of resonator's frequency.
The optical frequency measurement of the laser using the frequency
comb is performed in parallel to the interrogation.

The determination of the Nd:YAG laser frequency and of the EOM drive
frequency at which the resonator mode is maximally excited, $f_{\mathrm{EOM,res}}$,
allows us to determine the absolute resonator frequency $f_{\mathrm{res}}=f_{\mathrm{laser,av}}\pm f_{\mathrm{EOM,res}}$.
The sign is chosen according to which of the two sidebands interrogates
the resonator.

\section{{\normalsize{}Results and Discussion\label{sec:Results-and-Discussion}}}

\subsection{{\normalsize{}Transmission signal measurement}}

An example of a frequency scan over the resonance is shown in Fig.~\ref{fig:LorFit}.
A Lorentzian fit yields the sideband frequency for achieving resonance,
$f_{\mathrm{EOM,res}}$. 
\begin{center}
\begin{figure}[H]
\begin{centering}
\includegraphics[scale=0.1]{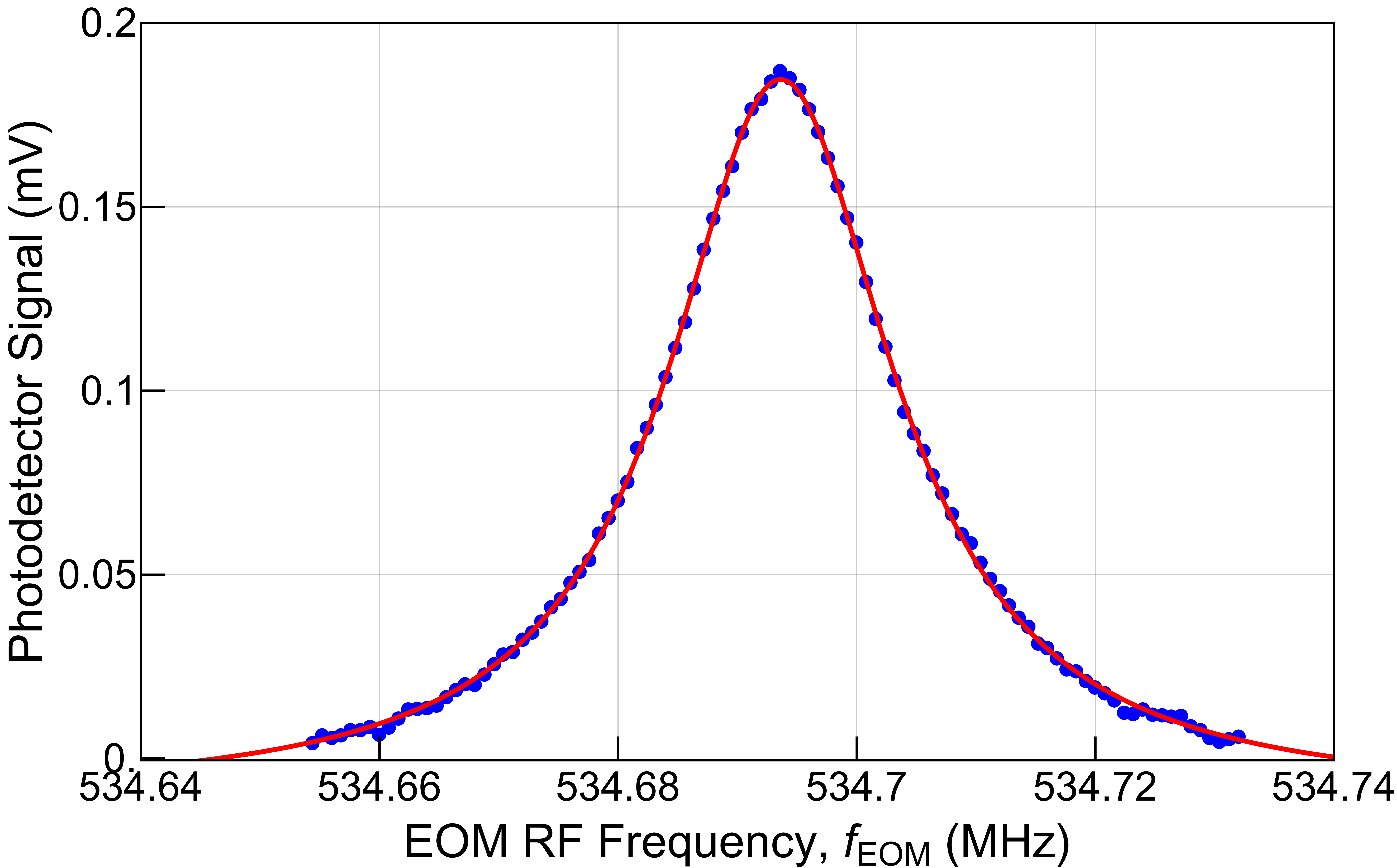}
\par\end{centering}
\caption{\label{fig:LorFit}A resonator transmission signal recorded during
a laser frequency scan. Blue: the transmission signal measured with
PD. Red: Lorentzian fit to the data.}
\end{figure}
\par\end{center}

\subsection{{\normalsize{}Characterization of thermal expansion}}

In order to achieve an accurate characterization of the resonator's
long-term frequency drift, we first determine its \textcolor{black}{thermal
time constant}. The resonator was initially set to $20\,^{\circ}\mathrm{C}$.
After the set temperature was changed to $19\,^{\circ}\mathrm{C}$,
the resonator's frequency was repeatedly measured at intervals of
10~minutes. The result is shown in Fig.~\ref{fig:cteTimeConstant}.
The thermal time constant is determined to be $2\times10^{4}$~s
$\approx$~5.6~hours, by fitting an exponential decay curve to the
data.
\begin{center}
\begin{figure}[H]
\centering{}\includegraphics[scale=0.1]{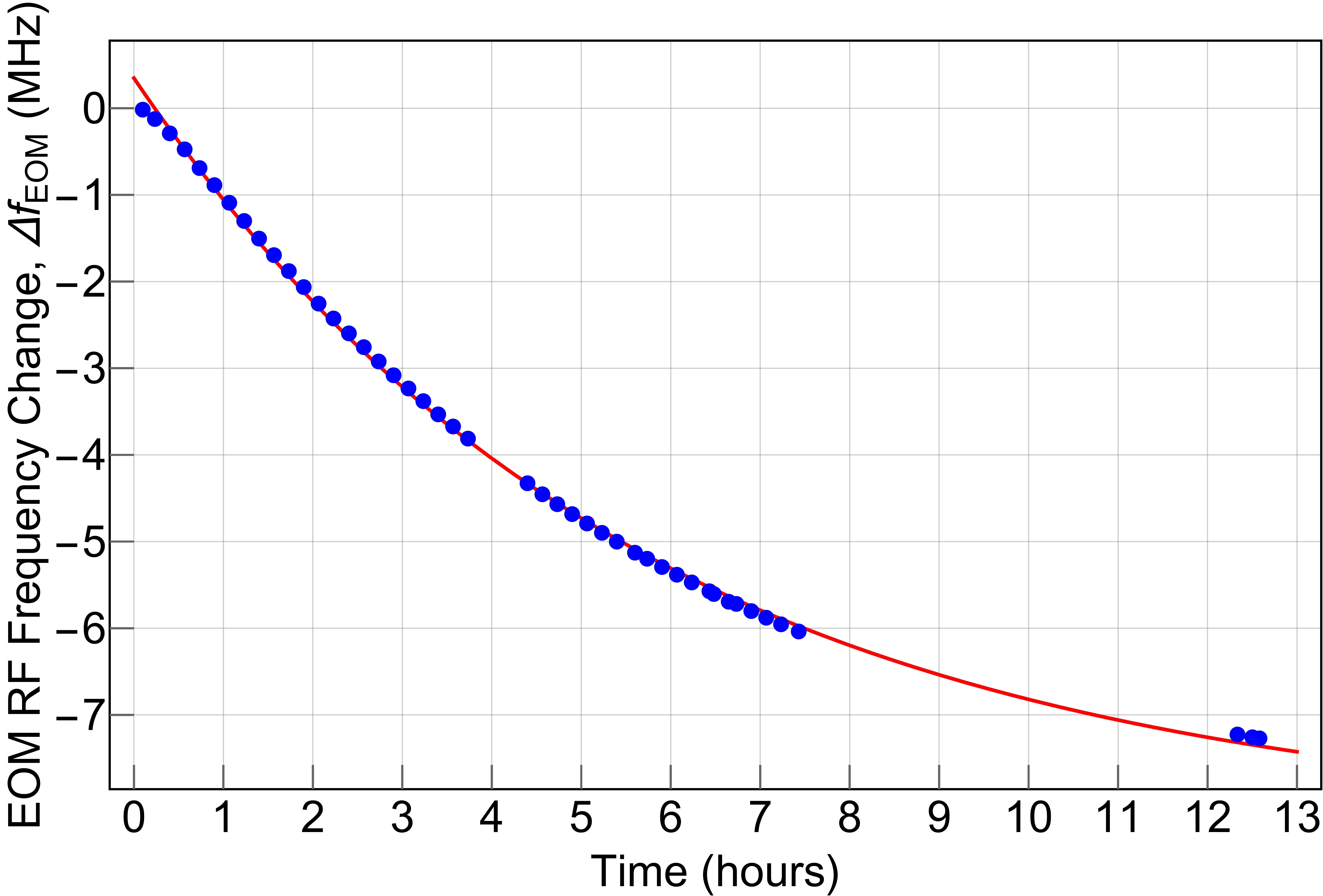}\caption{\label{fig:cteTimeConstant}Determination of thermal time constant
from the reaction of the resonator frequency to a 1~K temperature
change of the set point. The EOM RF frequency change, $\Delta f_{\mathrm{EOM}}=f_{\mathrm{EOM}}-f_{\mathrm{EOM},0}$.
Blue: data. Red: exponential decay fit.}
\end{figure}
\par\end{center}

Since the thermal settling follows an exponential decay, we decided
to perform measurements always 8 hours after the temperature change
to allow for two measurements per workday and find the zero-CTE point
within a reasonable time. The resonator frequencies $f_{\mathrm{res}}(T)$
at temperatures $T$ between $16\,^{\circ}\mathrm{C}$ and $29\,^{\circ}\mathrm{C}$
were measured at intervals of $1\,\mathrm{K}$ or smaller, in order
to obtain a precise determination of the zero-CTE temperature. Here,
measurement of only $f_{\mathrm{EOM}}$ is sufficient because the
change in the resonator's frequency due to change in temperature is
much larger than any drifts in the laser frequency. Fig.~\ref{fig:zerocte}
shows the data. A quadratic fit to the data was performed, and the
zero-CTE temperature was determined from the turning point of the
fit function. 
\begin{center}
\begin{figure}[h]
\centering{}\includegraphics[scale=0.1]{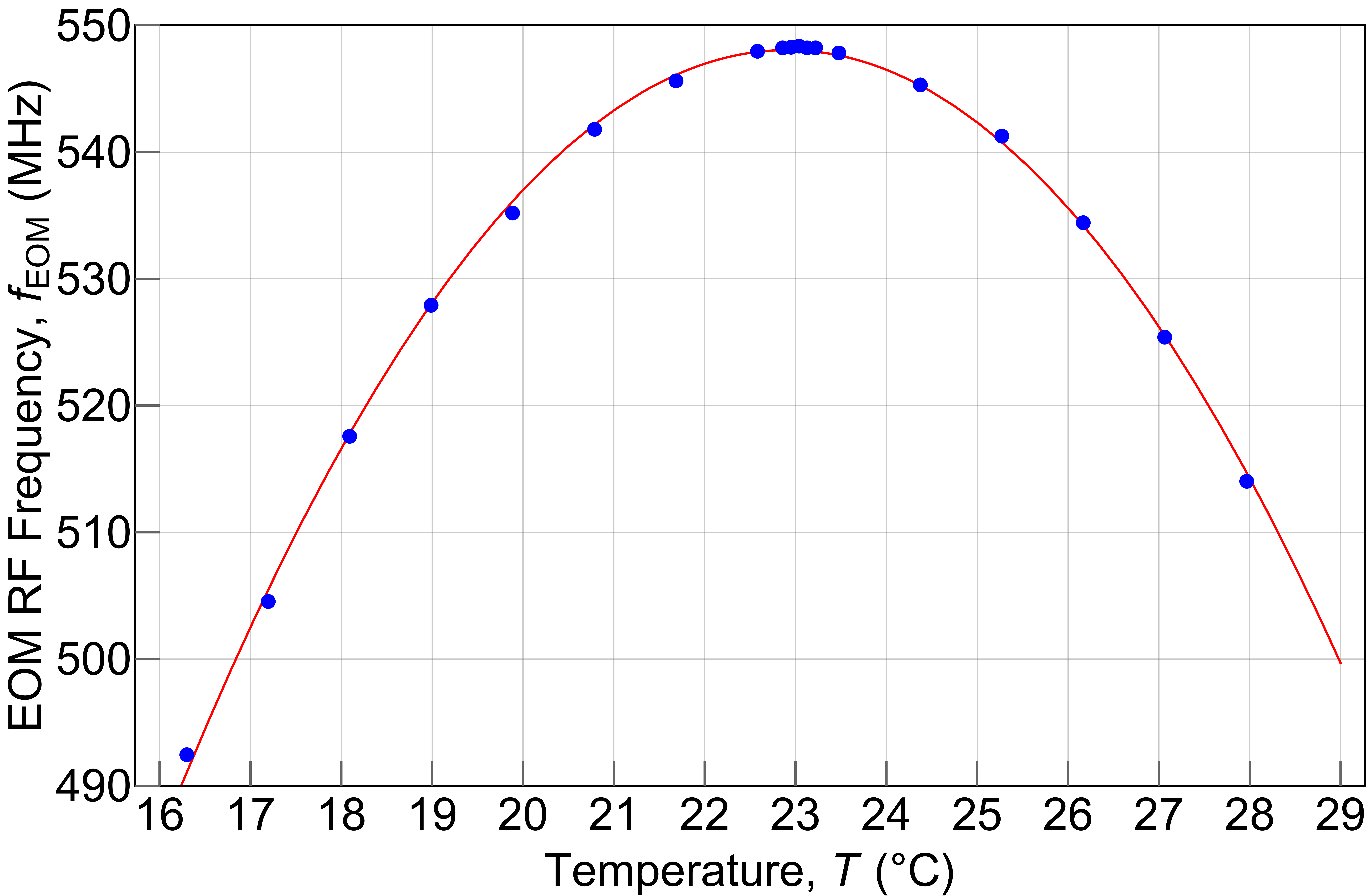}\caption{\label{fig:zerocte}NEXCERA resonator frequency at temperatures between
$16\,^{\circ}\mathrm{C}$ and $29\,^{\circ}\mathrm{C}$. Blue: measurement
data. Red: a quadratic fit to the data. }
\end{figure}
\par\end{center}

The zero-CTE temperature is found to be at $T_{0}=22.9\pm0.2\,{}^{\circ}\mathrm{C}$.
Both our measurement and the result by Hosaka \emph{et. al.} \citep{Hosaka2013},
$16.4\,^{\circ}\mathrm{C}$, yield values near room temperature.
\begin{center}
\begin{figure}[H]
\begin{centering}
\includegraphics[scale=0.34]{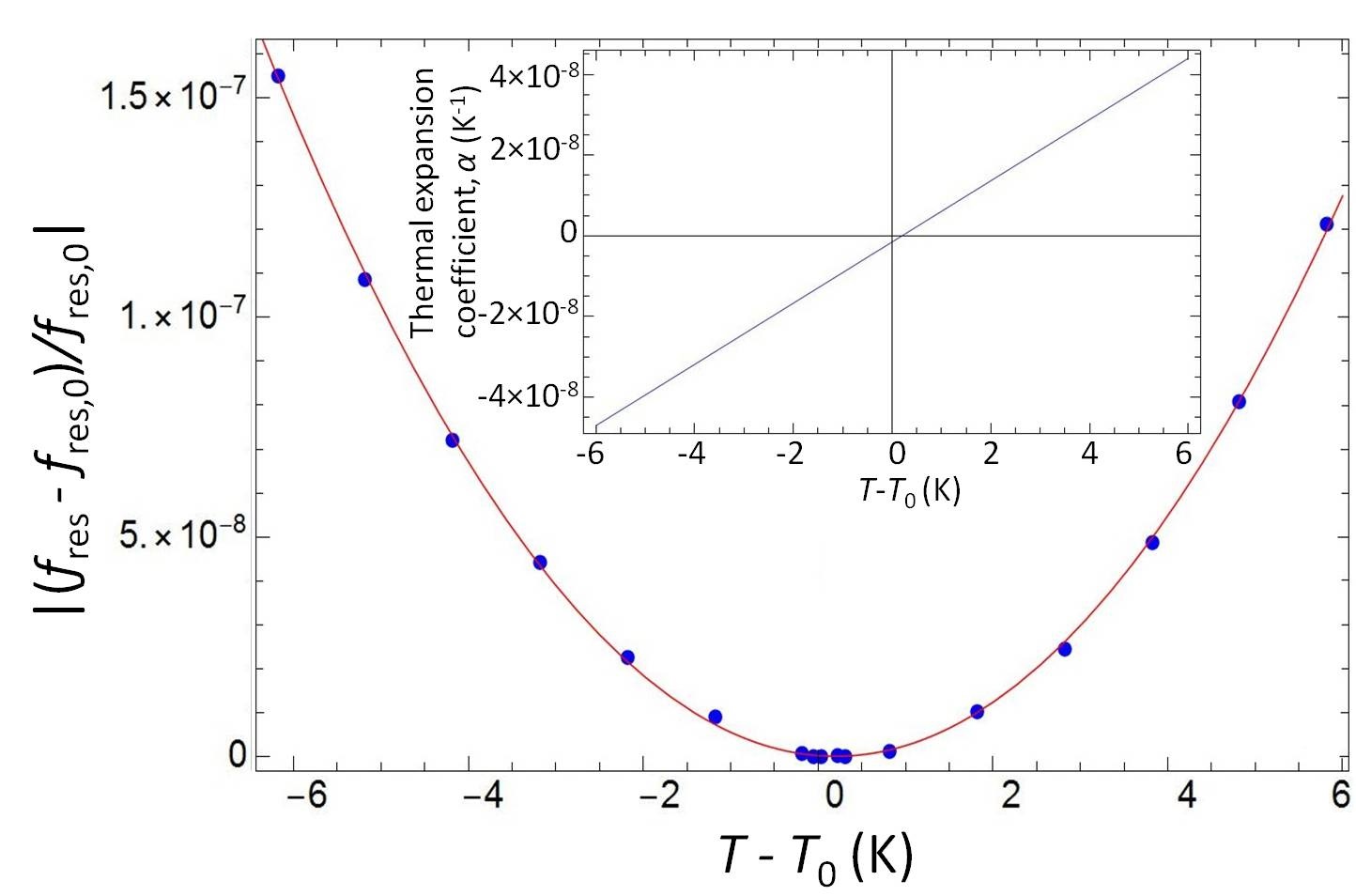}
\par\end{centering}
\caption{\label{fig:ctePlots}Normalised resonator frequency change around
its zero-CTE temperature, $T_{\mathrm{0}}$. Blue: experimental data.
Red: quadratic fit to data. Inset: thermal expansion coefficien\textcolor{black}{t,}\textcolor{red}{{}
${\color{black}\alpha}$} as a function of temperature.}
\end{figure}
\par\end{center}

From Fig.~\ref{fig:ctePlots}, a quadratic function is fitted to
the experimental data. The thermal expansion coefficient, $\alpha$,
i.e. the derivative of the fitted function is plotted inset. Our CTE
temperature derivative, $d\alpha(T_{0})/dT=$($7.58\pm0.08$)$\times10^{-9}$~$\mathrm{K}^{-2}$,
is approximately 2 times larger than the value of Hosaka et al., ($3.86\pm0.03$)$\times10^{-9}$~$\mathrm{K}^{-2}$. 

The NEXCERA resonator temperature was maintained at its zero-CTE temperature
value throughout the characterization of its long-term frequency drift.

\subsection{{\normalsize{}Absolute frequency measurements}}

\textcolor{black}{For the characterization of long-term frequency
drift, the cavity interrogations and Nd:YAG laser frequency measurements
were performed once per weekday. Laser frequency measurements and
sideband frequency $f_{\mathrm{EOM,res}}$ were both averaged over
approximately 30~minutes, and the resonator frequency $f_{\mathrm{res}}$
was computed from these time-averaged frequencies. The frequency change
was then given by the difference between the resonator frequency relative
to the initial frequency $f_{\mathrm{res}}(0)$, i.e. $\Delta f_{\mathrm{res}}(t)=f_{\mathrm{res}}(t)-f_{\mathrm{res}}(0)$.
The initial frequency measurement was performed on 25~April 2017,
125~days after optical contacting.}

\textcolor{black}{Fig.~\ref{fig:longtermDrift} shows the resonator
frequency change plotted against time. The frequency measurement was
carried out over a period of approximately 1608 hours, i.e. 67 days,
as shown in Fig.~\ref{fig:longtermDrift}. The data points, plotted
in green, exhibit a temporary offset of 125~kHz which could be due
to external mechanical perturbations on the experimental setup.}
\begin{center}
\begin{figure}[H]
\centering{}\includegraphics[scale=0.1]{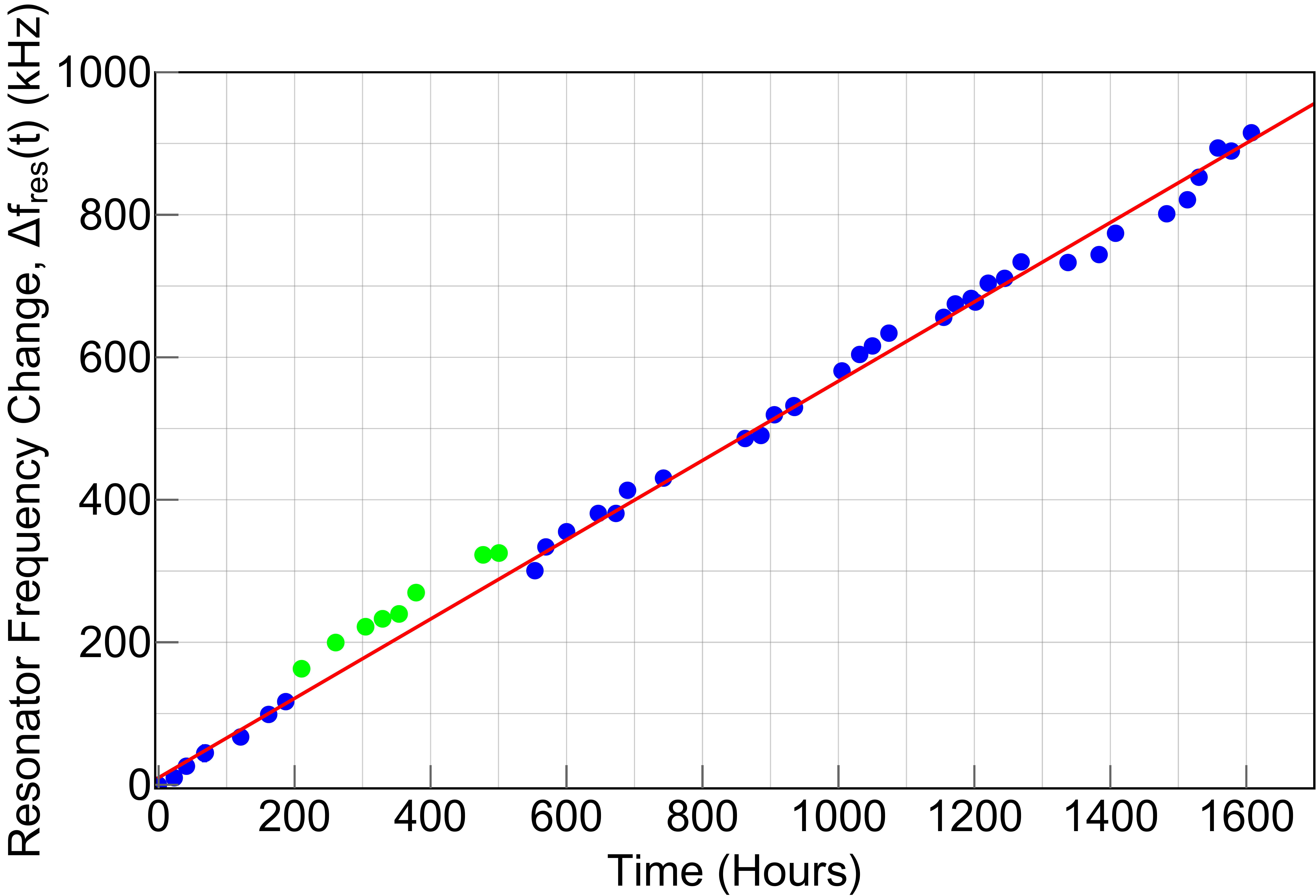}\caption{\label{fig:longtermDrift} \textcolor{black}{Long-term frequency drift
of the NEXCERA resonator. The red line is a linear fit to the blue
data points. The green data points are ignored.} }
\end{figure}
\par\end{center}

\textcolor{black}{From linear fit to the data, we find $d\Delta f_{\mathrm{res}}/dt=(0.155\pm0.001)$~Hz/s.
The fit to the data was done excluding the green data points. These
rates correspond to fractional frequency drift rates $f_{\mathrm{res}}(0)^{-1}d\Delta f_{\mathrm{res}}/dt=(5.50\pm0.04)\times10^{-16}$/s.
The error is statistical $1\sigma$ standard error; systematic errors
due to the frequency measurement technique are negligible. }

\section{{\normalsize{}Discussion and conclusion\label{sec:Conclusion}}}

\textcolor{black}{We have observed, and precisely measured, the drift
of the frequency of a resonator comprising a NEXCERA spacer and a
pair of high finesse ULE mirrors. We have found a nonzero drift rate.
If we neglect a possible contribution of the optical contacts between
the ULE mirror substrates and the NEXCERA spacer, we can assign this
frequency drift to an equal but opposite length drift rate $L^{-1}dL/dt=-f_{\mathrm{res}}(0)^{-1}d\Delta f_{\mathrm{res}}/dt$.
Thus, we find a length contraction, $L^{-1}d\Delta L/dt={\color{black}{\color{red}{\color{black}\left(-1.74\pm0.01\right)}}}\times10^{-8}$~$\mathrm{yr}^{-1}$
(interval II). This rate is of the same sign and comparable in magnitude
to that of ULE resonators. Indeed, the drift rate of our own ULE-resonator
stabilized Nd:YAG laser (see Fig.~\ref{fig:The-optical-setup}) amounted
to $f_{\mathrm{laser}}(0)^{-1}d\Delta f_{\mathrm{laser}}/dt=(0.625\pm0.003)\times10^{-8}$~$\mathrm{yr}^{-1}$.
The error of our NEXCERA drift rate measurement is approximately $0.01\times10^{-8}\,{\rm yr^{-1}}$,
50 times lower compared to the measurement in ref.~\citep{Takahashi2012}.
To ensure that the long-term frequency drift measurement was not affected
by the quality of the optical contacts, we tested them visually and
mechanically after completion of the measurements. No conspicuous
features were found.}

\textcolor{black}{Upon the completion of this work, an indepedent
study on NEXCERA N117B has been published \citep{Ito2017}, where
a significantly lower drift, 4.9 mHz/s was measured at the wavelength
of 1064~nm. The NEXCERA N118C, used in this work, and the N117B have
similar chemical compositions and microstructures but are sintered
in different atmospheres, i.e. the N117B is sintered in Argon atmosphere
while N118C is sintered in air. In view of the differences in long-term
drift between the materials, it is of interest to characterize more
samples of these materials, in order to determine whether the observed
drift rate difference is indeed a reproducible property.}

\textcolor{black}{Despite the difference in the measurement results
of the long-term drifts for optical resonators made from different
type of NEXCERA ceramic, NEXCERA, nevertheless, remains a promising
material for applications in the field of laser frequency stabilization.
For such applications, first proposed by Hosaka et al. \citep{Hosaka2013},
NEXCERA has the potential advantages of larger ratio of Young's modulus
to density, hence potentially lower acceleration sensitivity than
ULE. Also, it can be produced in larger sizes (up to 1~m), potentially
opening an avenue to lower thermal noise by increasing the resonator's
length. The larger CTE temperature derivative compared to ULE can
be compensated by advanced passive and active temperature stabilization. }

We are indebted to A. Nevsky for help with vacuum chamber as well
as helpful discussions on this work, U.~Rosowski for help with the
frequency comb measurement and \linkToOrcidNameIcon{E.~Wiens}{https://orcid.org/0000-0002-0013-1727}
for help with the resonator scans. We thank D.~Iwaschko also for
his support in providing the electronics of the experimental setup.
We acknowledge useful discussions with D.~V.~Sutyrin, E.~Magoulakis,
\linkToOrcidNameIcon{S.~Alighanbari}{https://orcid.org/0000-0003-0111-9794}
and M.~Schioppo. This work was funded by European Union grants FP7-PEOPLE-2013-ITN
Number 607491 ``COMIQ'' and FP7-PEOPLE-2013-ITN Number 607493 ``FACT''.

%\bibliographystyle{plainnat}
%\bibliography{all-references}

\begin{thebibliography}{15}
\providecommand{\natexlab}[1]{#1}
\providecommand{\url}[1]{\texttt{#1}}
\expandafter\ifx\csname urlstyle\endcsname\relax
  \providecommand{\doi}[1]{doi: #1}\else
  \providecommand{\doi}{doi: \begingroup \urlstyle{rm}\Url}\fi

\bibitem[Sch(2011)]{Schott2}
{Manufacturer's Specification, Zerodur, Schott AG }.
\newblock 2011.
\newblock URL
  \url{http://www.schott.com/d/advanced\_optics/f7ae3c11-0226-4808-90c7-59d6c8816daf/1.0/schott\_zerodur
  \_katalog\_july\_2011\_en.pdf}.

\bibitem[Cor(2016)]{Corning2}
{Manufacturer's Specification, ULE, Corning Inc.}
\newblock 2016.
\newblock URL
  \url{https://www.corning.com/media/worldwide/csm/documents/7972%20ULE%20Product%20Information%20Jan%202016.pdf}.

\bibitem[Alnis et~al.(2008)Alnis, Matveev, Kolachevsky, Udem, and
  H\"ansch]{Alnis2008}
J.~Alnis, A.~Matveev, N.~Kolachevsky, Th. Udem, and T.~W. H\"ansch.
\newblock Subhertz linewidth diode lasers by stabilization to vibrationally and
  thermally compensated ultralow-expansion glass {Fabry-P\'erot} cavities.
\newblock \emph{Phys. Rev. A}, \textbf{77}:\penalty0 053809, 2008.
\newblock \doi{10.1103/PhysRevA.77.053809}.

\bibitem[Berthold et~al.(1977)Berthold, Jacobs, and Norton]{Berthold1977}
J.~W. Berthold, S.~F. Jacobs, and M.~A. Norton.
\newblock Dimensional stability of fused silica, invar, and several ultra-low
  thermal expansion materials.
\newblock \emph{Metrologia}, \textbf{13}\penalty0 (1):\penalty0 9, 1977.
\newblock \doi{10.1088/0026-1394/13/1/004}.

\bibitem[Chen et~al.(2014)Chen, Nevsky, Cardace, Schiller, Legero, H{\"a}fner,
  Uhde, and Sterr]{Chen2014}
Q.-F. Chen, A.~Nevsky, M.~Cardace, S.~Schiller, T.~Legero, S.~H{\"a}fner,
  A.~Uhde, and U.~Sterr.
\newblock A compact, robust, and transportable ultra-stable laser with a
  fractional frequency instability of $1 \times 10^{-15}$.
\newblock \emph{Review of Scientific Instruments}, \textbf{85}\penalty0
  (11):\penalty0 113107, 2014.
\newblock \doi{10.1063/1.4898334}.

\bibitem[H\"{a}fner et~al.(2015)H\"{a}fner, Falke, Grebing, Vogt, Legero,
  Merimaa, Lisdat, and Sterr]{Hafner2015}
S.~H\"{a}fner, S.~Falke, C.~Grebing, S.~Vogt, T.~Legero, M.~Merimaa, C.~Lisdat,
  and U.~Sterr.
\newblock $8\times10^{-17}$ fractional laser frequency instability with a long
  room-temperature cavity.
\newblock \emph{Opt. Lett.}, \textbf{40}\penalty0 (9):\penalty0 2112--2115,
  2015.
\newblock \doi{10.1364/OL.40.002112}.

\bibitem[Hosaka et~al.(2013)Hosaka, Inaba, Akamatsu, Yasuda, Sugawara, Onae,
  and Hong]{Hosaka2013}
K.~Hosaka, H.~Inaba, D.~Akamatsu, M.~Yasuda, J.~Sugawara, A.~Onae, and F.-L.
  Hong.
\newblock A {Fabry-P{\'e}rot} etalon with an ultralow expansion ceramic spacer.
\newblock \emph{Japanese Journal of Applied Physics}, \textbf{52}\penalty0
  (3R):\penalty0 032402, 2013.

\bibitem[Ito et~al.(2017)Ito, Silva, Nakamura, and Kobayashi]{Ito2017}
Isao Ito, Alissa Silva, Takuma Nakamura, and Yohei Kobayashi.
\newblock Stable {CW} laser based on low thermal expansion ceramic cavity with
  4.9 {mHz/s} frequency drift.
\newblock \emph{Opt. Express}, 25\penalty0 (21):\penalty0 26020--26028, 2017.
\newblock \doi{10.1364/OE.25.026020}.

\bibitem[Jacobs and Bass(1989)]{Jacobs1989}
S.~F. Jacobs and D.~Bass.
\newblock {Improved dimensional stability of Corning 9600 and Schott Zerodur
  glass ceramics}.
\newblock \emph{Appl. Opt.}, \textbf{28}\penalty0 (19):\penalty0 4045--4046,
  1989.
\newblock \doi{10.1364/AO.28.004045}.

\bibitem[Keller et~al.(2014)Keller, Ignatovich, Webster, and
  Mehlst{\"a}ubler]{Keller2014}
J.~Keller, S.~Ignatovich, S.~A. Webster, and T.~E. Mehlst{\"a}ubler.
\newblock Simple vibration-insensitive cavity for laser stabilization at the
  $10^{-16}$ level.
\newblock \emph{Applied Physics B}, \textbf{116}\penalty0 (1):\penalty0
  203--210, 2014.
\newblock ISSN 1432-0649.
\newblock \doi{10.1007/s00340-013-5676-y}.

\bibitem[Keupp et~al.(2005)Keupp, Douillet, Mehlst{\"a}ubler, Rehbein, Rasel,
  and Ertmer]{Keupp2005}
J.~Keupp, A.~Douillet, T.~E. Mehlst{\"a}ubler, N.~Rehbein, E.~M. Rasel, and
  W.~Ertmer.
\newblock A high-resolution {Ramsey-Bord{\'e}} spectrometer for optical clocks
  based on cold {Mg} atoms.
\newblock \emph{The European Physical Journal D - Atomic, Molecular, Optical
  and Plasma Physics}, \textbf{36}\penalty0 (3):\penalty0 289--294, 2005.
\newblock ISSN 1434-6079.
\newblock \doi{10.1140/epjd/e2005-00302-7}.

\bibitem[Marmet et~al.(1997)Marmet, Madej, Siemsen, Bernard, and
  Whitford]{Marmet1997}
L.~Marmet, A.~A. Madej, K.~J. Siemsen, J.~E. Bernard, and B.~G. Whitford.
\newblock Precision frequency measurement of the {$^2\mathrm{S}_{1/2} -
  ^2\mathrm{D}_{5/2}$} transition of $\mathrm{Sr}^+$ with a 674-nm diode laser
  locked to an ultrastable cavity.
\newblock \emph{IEEE Transactions on Instrumentation and Measurement},
  \textbf{46}\penalty0 (2):\penalty0 169--173, 1997.
\newblock ISSN 0018-9456.
\newblock \doi{10.1109/19.571804}.

\bibitem[Takahashi(2010)]{Takahashi2010}
A.~Takahashi.
\newblock Long-term dimensional stability and longitudinal uniformity of line
  scales made of glass ceramics.
\newblock \emph{Measurement Science and Technology}, \textbf{21}\penalty0
  (10):\penalty0 105301, 2010.
\newblock \doi{10.1088/0957-0233/21/10/105301}.

\bibitem[Takahashi(2012)]{Takahashi2012}
A.~Takahashi.
\newblock Long-term dimensional stability of a line scale made of low thermal
  expansion ceramic {NEXCERA}.
\newblock \emph{Measurement Science and Technology}, \textbf{23}\penalty0
  (3):\penalty0 035001, 2012.
\newblock \doi{10.1088/0957-0233/23/3/035001}.

\bibitem[Zhu and Hall(1992)]{Zhu1992}
M.~Zhu and J.~L. Hall.
\newblock Short and long term stability of optical oscillators.
\newblock In \emph{Proceedings of the 1992 IEEE Frequency Control Symposium},
  pages 44--55, 1992.
\newblock \doi{10.1109/FREQ.1992.270036}.

\end{thebibliography}

\end{document}